# Orientation of molecules via laser-induced anti-alignment

E. Gershnabel , I. Sh. Averbukh
*Dept. of Chemical Physics, The Weizmann Institute of Science, Rehovot 76100, ISRAEL*
R.J. Gordon
*Dept. of Chemistry , University of Illinois at Chicago, Chicago, IL 60680-7061, USA*

## Abstract

We show that field-free molecular orientation induced by a half-cycle pulse may be considerably enhanced by an additional laser pulse inducing molecular anti-alignment. Two qualitatively different enhancement mechanisms are identified depending on the pulse order, and their effects are optimized with the help of quasi-classical as well as fully quantum models.

The orientation and alignment of molecules have long been of interest in chemistry and physics. Alignment and anti-alignment conventionally refer to head-on vs. broadside localization of the symmetry axis of a molecule, whereas orientation refers to control of the up and down directions of an aligned molecule. Modern applications of aligned and oriented molecules, such as high harmonic generation [1], laser pulse compression [2], nanolithography [3], control of photodissociation and photoionization processes[4], and quantum information processing[5], have motivated the development of all-optical techniques for aligning molecules under field-free conditions. A major advance has been the use of linearly polarized, ultrashort laser pulses to create a rotational wave packet by an impulsive Raman mechanism. If the laser pulse is much shorter than the rotational period of the molecule (i.e., if its bandwidth is greater than the rotational level spacing), the molecule undergoes a series of Raman excitations that produce a coherent superposition of rotational states[6].

For short pulses, peak field-free alignment along the electric vector of the laser field is achieved after termination of the laser pulse, at a time that depends on the pulse strength. As the wave packet evolves, the molecule loses its alignment, and even becomes anti-aligned some later time. The molecule also undergoes a series of field-free realignments [7] at integer multiples of the revival time, $\tau_{rev} = \pi\hbar / B$, where $B = \hbar^2 /(2I_m)$ is the rotational constant ($I_m$ is the moment of inertia). In addition, a number of fractional rotational revivals occur at rational fractions of $\tau_{rev}$ [8].

Orientation of molecules is technically more difficult to achieve than alignment because it requires an asymmetric field that distinguishes between up and down. A variety of methods have been proposed to break the field symmetry, including introduction of a weak DC electric (or magnetic) field in conjunction with a pulsed laser field[9], and coherent excitation with laser fields of frequencies ω and 2ω[10]. The most versatile proposal to orientation of dipolar molecules utilizes asymmetric electromagnetic half-cycle pulses (HCP)[11,12]. Similar to the case of laser-induced alignment, maximal orientation is achieved under field-free conditions after the end of the HCP. As was shown in Ref. [12], a single short HCP cannot provide perfect orientation, and its effect saturates with intensity. Because of the random initial direction of the molecular dipole, different molecules acquire different angular speeds after being kicked by a HCP. Molecules initially located at obtuse angles with respect to the HCP ($\pi/2 \le \theta \le \pi$) are too slow to catch up with molecules starting from acute angles ($0 \le \theta \le \pi/2$). As a result, the kicked molecules do not all point in the same final direction, and the orientation factor has a maximum value of $<\cos(\theta)>\approx 0.75$. This effect is similar to non-perfect focusing caused by spherical aberration in geometrical optics. It was shown theoretically[12, 13, 14] that the degree of field-free orientation and alignment could be enhanced using trains of laser pulses. Moreover, strong alignment by a pair of pulses has been experimentally achieved[15,16]. Orienting molecules with multiple HCP's currently is a challenging experimental task because of the difficulty in making controlled pulse trains of sufficient strength.

In this Letter we propose a method of substantially increasing the degree of orientation by combining an asymmetric half-cycle pulse with a symmetric femtosecond laser pulse inducing *anti-alignment* in a molecular ensemble. Depending on the temporal order of pulses, we have identified two qualitatively different mechanisms for the orientation enhancement, which we term "orienting an anti-aligned state" and "correcting the rotational velocity aberration", respectively. In the former mechanism, a symmetric laser pulse pushes the molecular symmetry axis toward the plane perpendicular to the desired orientation direction and prepares the molecules in an *anti-aligned state* that is angularly localized near $\theta=\pi/2$. When a delayed strong asymmetric HCP is applied to such an ensemble, all the molecules gain nearly the same rotational velocity. Because all the molecules depart from the anti-aligned state with close initial angles, they reach the orientation axis almost simultaneously at some later time. We show that such a pulse pair is capable of producing much greater orientation than is possible with a single, arbitrarily strong HCP. In the second mechanism, the *anti-aligning* pulse is applied shortly *after* the orienting HCP (or even simultaneously with it). Such a pulse decelerates the rotation of molecular dipoles having an acute angle with respect to the orientation direction and accelerates dipoles having obtuse angles. This effect compensates for the "spherical aberration" in the angular distribution of the rotational velocity and improves the overall orientation at some later time. We demonstrate that significantly enhanced orientation can be achieved by a proper choice of the delay between pulses and of their relative intensities.

The theory of alignment and orientation of molecules with time-varying electric fields has been discussed in detail in previous publications (for a recent review, see[17]), and only the essential points are presented here. The Hamiltonian for a 3D driven rigid rotator (linear molecule) interacting with a linearly polarized field is given by

$$H(\theta,t)=\frac{\hat{J}^2}{2I_m}+V(\theta,t) \qquad (1)$$

where $\hat{J}$ is the angular momentum operator and $\theta$ is the angle between the molecular axis and the polarization vector of the field. For a symmetric laser pulse interacting with the induced polarization, the interaction term, averaged over the fast optical oscillations, is given by $V(\theta,t)=-(1/4)\,\varepsilon^2(t)\left[\left(\alpha_{//}-\alpha_{\perp}\right)\cos^2(\theta)+\alpha_{\perp}\right]$. Here, $\varepsilon(t)$ is the envelope of the laser field, and $\alpha_{//}$ and $\alpha_{\perp}$ are the parallel and perpendicular components of the polarizability tensor, respectively. For a symmetric laser pulse the contribution from the permanent dipole averages to zero. For an asymmetric HCP, the interaction with the dipole moment is given by $V(\theta,t)=-\mu\varepsilon(t)\cos(\theta)$, where $\mu$ is the permanent dipole moment and $\varepsilon(t)$ is the amplitude of the HCP. In the present paper we assume that the duration of the laser pulse is much shorter than the dominant periods of the rotational wave packet, so that the excitation dynamics may be calculated in the impulsive limit. The impulse imparted to the rotator is characterized by dimensionless kick strength, *P*.

For an asymmetric pulse, $P$ is given by $P_a = (\mu / \hbar)\int_{-\infty}^{\infty} \varepsilon(t)\, dt$, where the integration is performed over the unidirectional part of the half-cycle pulse, and $P_s = (1/4\hbar)(\alpha_{//} - \alpha_{\perp})\int_{-\infty}^{\infty} \varepsilon^2(t) dt$ for a symmetric pulse. We start with the mechanism of "orienting an anti-aligned state" and consider a rotator excited first with a symmetric pulse of strength $P_s$ at $t = 0$ and then with an asymmetric pulse of strength $P_a$ at $t = t_1$. Henceforth the dimensionless time is measured in the units of $I_m / \hbar = \tau_{rev} /(2\pi)$. Considerable physical insight may be derived from the (semi)-classical treatment of the problem, which is valid for $P_s, P_a >> 1$. This is a natural approach to the problem of enhanced orientation involving highly excited rotational states. Classically, if the rotator is initially aligned at angle $\theta_0$, it will be found at the same angle just after the first kick but with angular velocity $-P_s \sin(2\theta_0)$, so that at some later time it will have an angle $\theta(t) = \theta_0 - P_s t \sin(2\theta_0)$. When at time $t = t_1$ the orienting pulse of strength $P_a$ is applied, the velocity increment is $-P_a \sin\theta(t_1)$, so that the angular velocity after the second pulse is

$$\omega(\theta_0) = -P_s \sin(2\theta_0) - P_a \sin\left[\theta_0 - P_s\, t_1 \sin(2\theta_0)\right]$$

Therefore, the angle $\theta$ at time $t_2$ after the second pulse is

$$\theta(t_1 + t_2) = \theta_0 - P_s\, t_1 \sin(2\theta_0) - t_2 \left\{P_s \sin(2\theta_0) + P_a \sin\left[\theta_0 - P_s\, t_1 \sin(2\theta_0)\right]\right\}. \qquad (2)$$

A similar expression may be derived for the inverse order of pulses, needed for the second mechanism of enhanced orientation. The orientation and alignment factors at $t = t_1 + t_2$ are calculated by averaging $\cos^k \theta(t)$ over all values of $\theta_0$,

$$\left\langle \cos^k \theta(t) \right\rangle = \frac{1}{2}\int_0^{\pi} \cos^k \theta(t) \sin\theta_0 \, d\theta_0, \tag{3}$$

where $k = 1$ and $2$ for orientation and alignment, respectively.

The same problem may be treated quantum-mechanically. A quantum rotator initially in the ground state at $t = 0$ acquires a phase factor produced by the first, short symmetric pulse, $\psi(\theta, t = +0) = \exp[iP_s \cos^2\theta]\, Y_0^0(\theta)$. By expanding this expression as a sum of spherical harmonics, one finds the wave function at later time $t$,

$$\psi(\theta, t) = \frac{1}{\sqrt{4\pi}} \sum_{J=0} c_J \exp[-iJ(2J+1)\, t]\, Y_{2J}^0(\theta). \tag{4}$$

The coefficients $c_J$ are given by[13]

$$c_J = \sqrt{\pi(4J+1)}(iP_s)^J \frac{\Gamma(J+1/2)}{\Gamma(2J+3/2)}\, {}_1F_1[J + 1/2, 2J + 3/2, iP_s]$$

where ${}_1F_1$ is the confluent hypergeometric function. At time $t = t_1$ the rotator is kicked by the orienting pulse, acquiring another phase factor, $\psi(\theta, t_1 + 0) = \exp[iP_a \cos\theta]\, \psi(\theta, t_1 - 0)$. We use the well-known expression

$$\exp(iP_a \cos\theta) = \sum_{j=0}^{\infty} i^j \sqrt{4\pi(2j+1)}\, j_j(P_a) Y_j^{\,0}(\theta)$$

where $j_j(P_a)$ is a spherical Bessel function, and again expand the wave function in a series of spherical harmonics, $\psi(\theta, t_1 + 0) = \sum_{l=0}^{\infty} d_l\, Y_l^0(\theta)$, where

$$d_l = \frac{1}{\sqrt{4\pi}} \sum_{l'=0}^{\infty} \sum_{j=0}^{\infty} i^j \sqrt{(2j+1)}\, j_j(P_a) c_{l'} \exp[-il'(2l'+1)\tau]$$
$$\times \sqrt{(2j+1)(4l'+1)(2l+1)} \frac{C(j, 2l', l \,|\, 0,0,0)^2}{2l+1}$$

Here $C(j,2l',l\,|\,0,0,0)$ is a Clebsch-Gordan coefficient. This new wave function is allowed to propagate freely until $t = t_1 + t_2$, at which point the orientation and alignment factors are calculated by $\langle \cos^k \theta \rangle = \langle \psi(\theta,t) | \cos^k \theta | \psi(\theta,t) \rangle$, $k = 1,2$.

Our analysis shows that strong transient anti-alignment may be achieved via two related methods. The first one is of a (quasi)-classical nature. It operates on a short time-scale ($t << \tau_{rev}$) and requires *negative* values of the kick strength $P_s$. Pulses with negative $P_s$ produce anti-alignment by pushing molecules into the equatorial plane ($\theta = \pi/2$). Figure 1a shows the expectation value of $<\cos^2\theta>$ calculated both classically and quantum-mechanically for a pulse with $P_s = -10$. Both treatments predict a deep minimum ($\langle \cos^2 \theta \rangle_{\min} = 0.077$) shortly after the pulse at $t_m \approx 0.8/|P_s|$. There are various ways of achieving a *negative* kick strength of the symmetric pulse. First, some alkali halides (such as LiF) have a negative polarizability anisotropy ($\alpha_{//} < \alpha_{\perp}$)[18]. Second, the interaction of molecules with a circularly polarized light pulse propagating in the direction of the desired orientation is proportional to $P_s \sin^2\theta = P_s - P_s \cos^2\theta$. This result is formally equivalent to the interaction with a linearly polarized pulse (our model) having negative kick strength. The second method of achieving anti-alignment uses laser pulses with *positive* $P_s$, which cause a substantial *alignment* on a short (classical) time scale *after* the kick (see Fig. 1b). However, if it were possible to invert the dynamics and travel back in time, one would observe a strong anti-alignment *before* the kick. Remarkably, the effect of quantum revivals[8] provides such an option. Indeed, $\psi(\theta, t_{rev} - \tau) = \psi(\theta, -\tau)$, and, for strong enough pulses, quantum dynamics just before one full revival cycle is equivalent to classical dynamics analytically continued to *negative times*. As a result, considerable anti-alignment is observed in this time domain (see Fig. 1b).

Figure 2 shows the results of direct numerical maximization of the orientation factor $|<\cos(\theta)>(P_s, P_a, t_1, t_2)|$ using the classical model (Eqs.(2) and (3)) with an anti-aligning pre-pulse ( $P_s < 0$ ). When optimizing, the model was formally extended to negative times to cover effects in the full revival time-domain. At zero initial temperature, the optimal solution depends only on $P_a / P_s$ and the products $P_s t_1$ and $P_a t_2$. Fig. 2 displays the highest post-pulse orientation factor $\langle\cos\theta\rangle$ (solid line) and the optimal delay between pulses ($t_{1opt}$) (dashed line) as a function of $P_a / |P_s|$. There are two optimal solutions of almost the same efficiency. The first one provides maximal orientation in the direction of the HCP shortly after the pulse (upper panel in Fig. 2). The second one (lower panel in Fig. 2) delivers enhanced orientation in the *opposite* direction in the full revival domain. In both cases, an impressive value of $|\langle\cos\theta\rangle_{\max}| \approx 0.95$ is achieved for rather modest pre-pulses ($|P_s| \sim 0.1\ P_a$) (as compared to the limit of $\langle\cos\theta\rangle_{\max} \approx 0.75$ for a single HCP). In this regime, the optimal delay between pulses indeed asymptotically approaches the moment of the best anti-alignment $t_m \approx 0.8 / |P_s|$, as was discussed in the introduction.

It is easy to show that the same degree of orientation may be achieved by combining a HCP with an *aligning* laser pulse ( $P_s > 0$ ). The apparent symmetry relations

$$<\cos(\theta)>(P_s,P_a,-t_1,-t_2)=<\cos(\theta)>(-P_s,-P_a,t_1,t_2) \quad (5)$$

$$<\cos(\theta)>(P_s,P_a,t_1,t_2)=-<\cos(\theta)>(P_s,-P_a,t_1,t_2) \quad (6)$$

reduce this problem to the already studied case of the anti-aligning pulse.

We used the same approach to analyze the second mechanism of enhanced orientation mentioned in the introduction. In the simplest, non-optimized version, the orienting and anti-aligning pulses are applied simultaneously ($t_1 = 0$). Direct numerical maximization of the expression in Eq.(3) shows that $<\cos(\theta)>_{\max}= 0.89$ when $P_a / |P_s| \approx 2.34$ and $|P_s| t_2 \approx 0.78$. When the hybrid pulse is composed of an orienting component and an *aligning* one ($P_s > 0$), the symmetry relations (5) and (6) predict the same orientation, but in the opposite direction ($<\cos(\theta)> = -0.89$) just before one full revival cycle ($P_s\, t_2 \approx -0.78$). This effect was observed in a recent paper[19] as a result of direct numerical simulation of the quantum rotational dynamics of molecules excited by a single hybrid pulse. More efficient results can be obtained when the HCP precedes the anti-aligning laser pulse. In this case, our (quasi)-classical analysis reveals a single dominating optimal solution presented in Fig. 3. The maximal orientation $<\cos(\theta)>_{\max} \approx 0.96$ is reached at $P_a / |P_s| \approx 1.6$ and the optimal delay is $t_1 \approx 0.36/|P_s|$.

We performed a fully quantum-mechanical analysis of the "orientation via anti-alignment" mechanisms using the above described methodology. Fig. 4 demonstrates the optimized values of the anti-aligning pulse strength, $|P_s|$, delay between pulses, $t_1$ and the maximal orientation factor, $\langle\cos\theta\rangle_{\max}$ as a function of $P_a$. Very good agreement between quantum and (quasi)-classical results is observed even for moderate anti-aligning and orienting pulses ( $P_s, P_a \sim 3$ ). Remarkably, a significantly enhanced orientation may be achieved with field strength available currently in the laboratory. Considering a KCl molecule in the ground state (revival time $t_{rev} \approx 128\, ps$ , dipole moment $\mu \approx 10.3\, D$, polarization anisotropy $(\alpha_{//} - \alpha_{\perp}) \approx 3.1\, A^3$, data taken from[18]) one expects $P_a \sim 10$ for a HCP with the amplitude of $100\, kV/cm$ and duration of about *2 ps* (1/*e* half width). According to Fig. 4, the orientation factor $<\cos(\theta)>_{\max} \approx 0.95$ may be observed if the HCP is followed by a delayed anti-aligning pulse of *2 ps* duration and $5 \times 10^{11}\, W/cm^2$ peak intensity. The orientation achievable with a given field strength is expected to decrease with temperature. Nevertheless, our preliminary calculations show that the hybrid scheme with delayed pulses produces robust enhanced orientation even at thermal conditions. Detailed results will be presented in a future publication.

The transparent physics behind the interplay between anti-alignment, orientation and revivals provides a solid basis for the future design of more sophisticated and efficient solutions. The hybrid scheme combining anti-aligning and orienting pulses may take advantage of other recent proposals for laser control of molecular alignment and orientation. Enhanced anti-alignment by trains of symmetric laser pulses may be performed prior to applying a HCP, similar to forced multi-pulse alignment techniques[12,13,14,15,16] . The "spherical aberrations" of a single HCP may be even better corrected by multiple series of delayed laser pulses. Finally, a hybrid pair of delayed pulses may be followed by a series of well-timed symmetric laser pulses designed to preserve the achieved orientation over an extended time period[20,21].

IA wishes to thank Israel Science Foundation for support of this research, and RJG acknowledges Motorola Corporation and the US Department of Energy for its support.

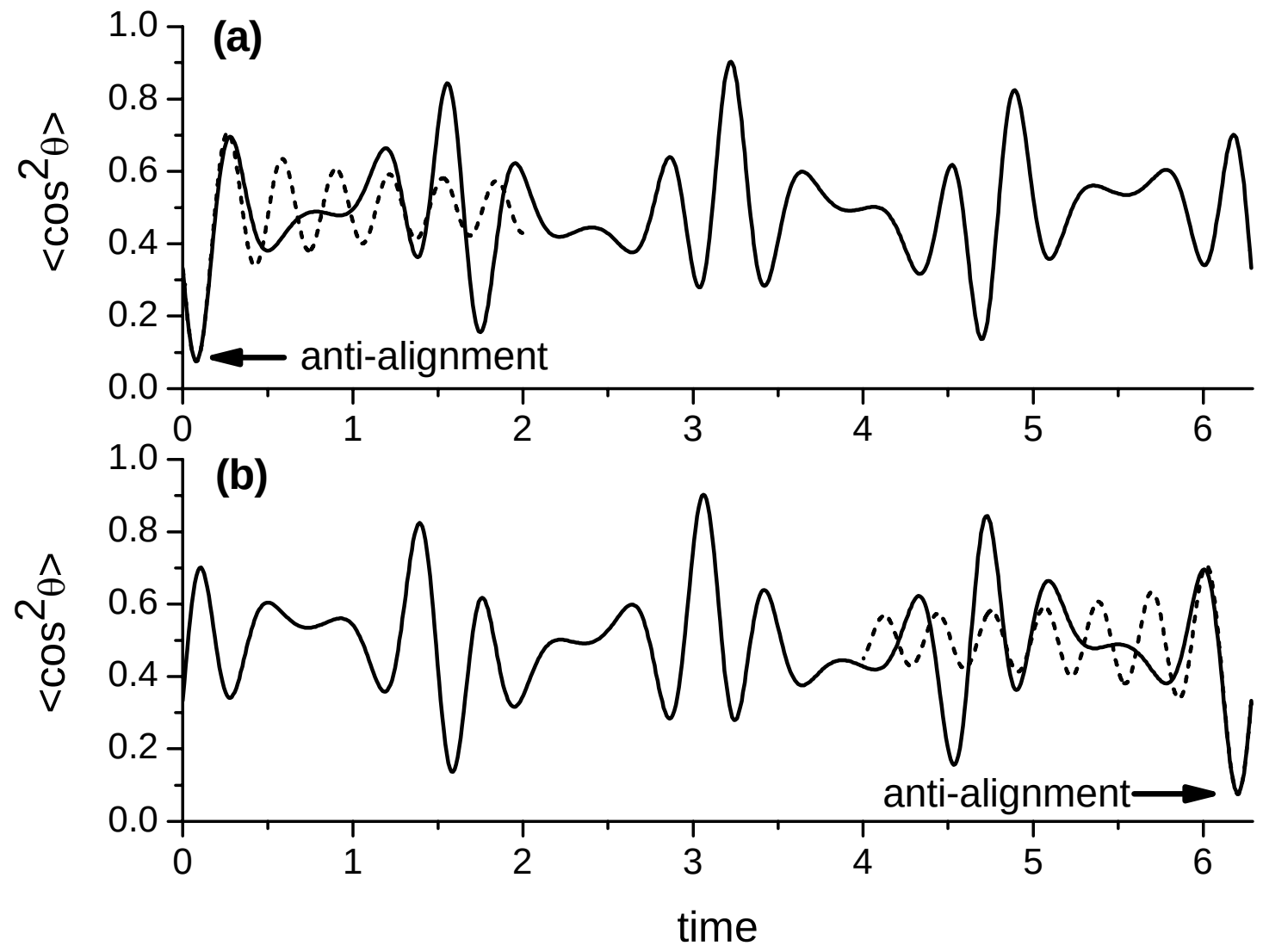

Figure 1. Alignment factor vs time after excitation by a laser pulse with (a) $P_s = -10$ and (b) $P_s = 10$ . Solid curves present quantum results. Dashed curves are calculated classically (a) for positive time and (b) for negative time (shifted by $\tau_{rev}$). Strong anti-alignment is seen in both cases.

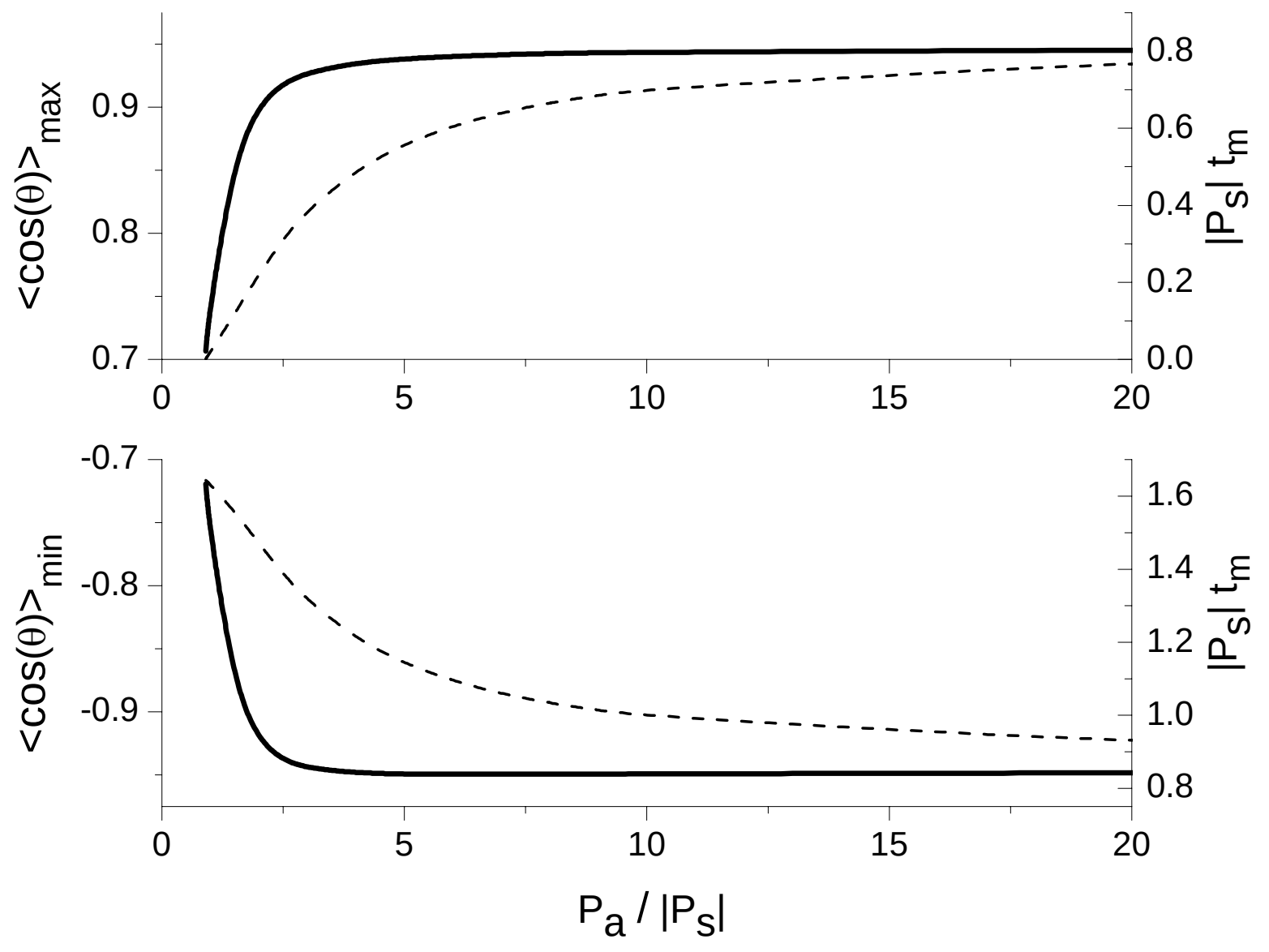

Figure 2. Classically optimized orientation factor (solid curves) and delay between pulses (dashed curves) . Laser pulse is fired before the half-cycle pulse.

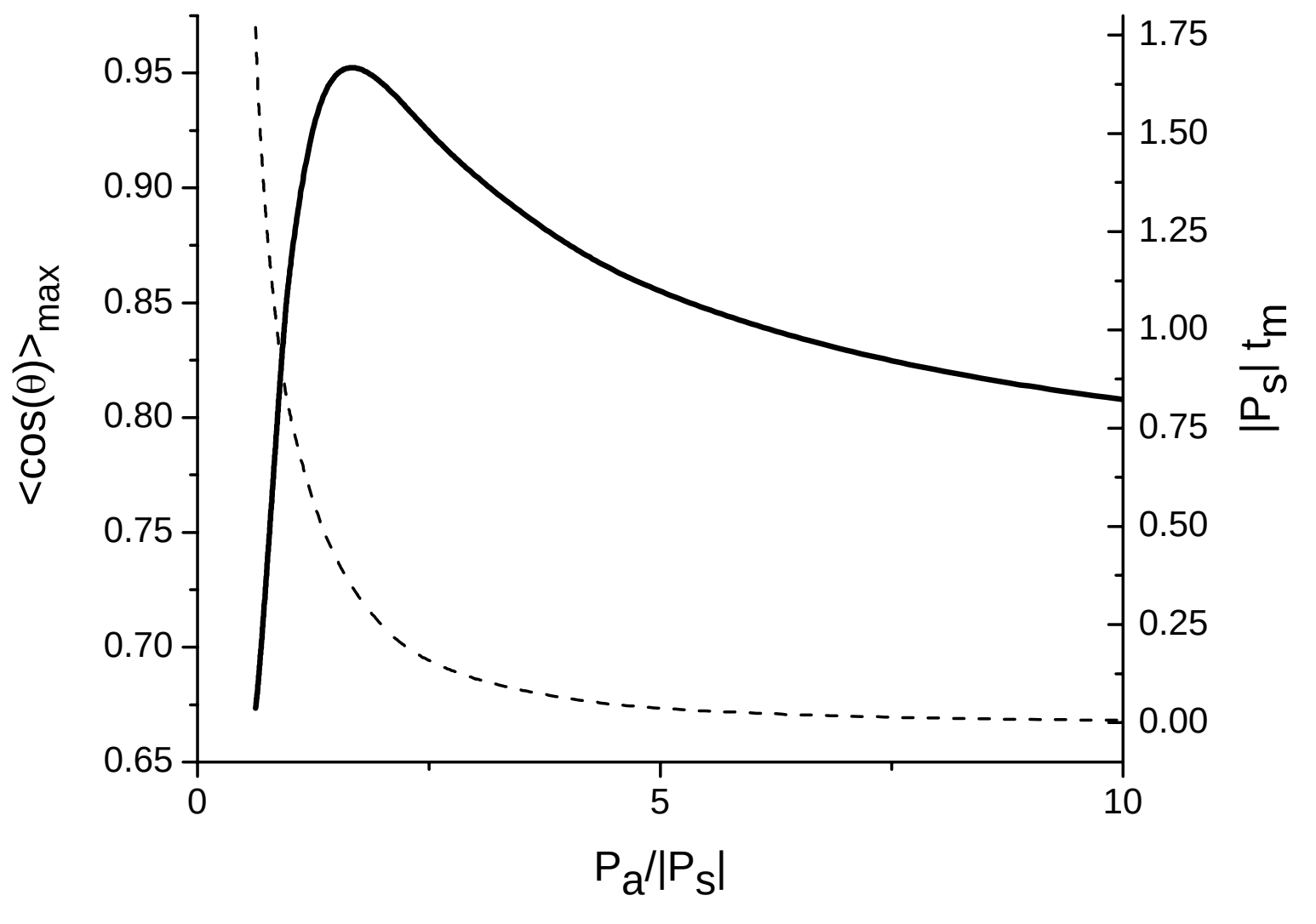

Figure 3. Classically optimized orientation factor (solid curve) and delay between pulses (dashed curve) . Laser pulse is fired after the half-cycle pulse.

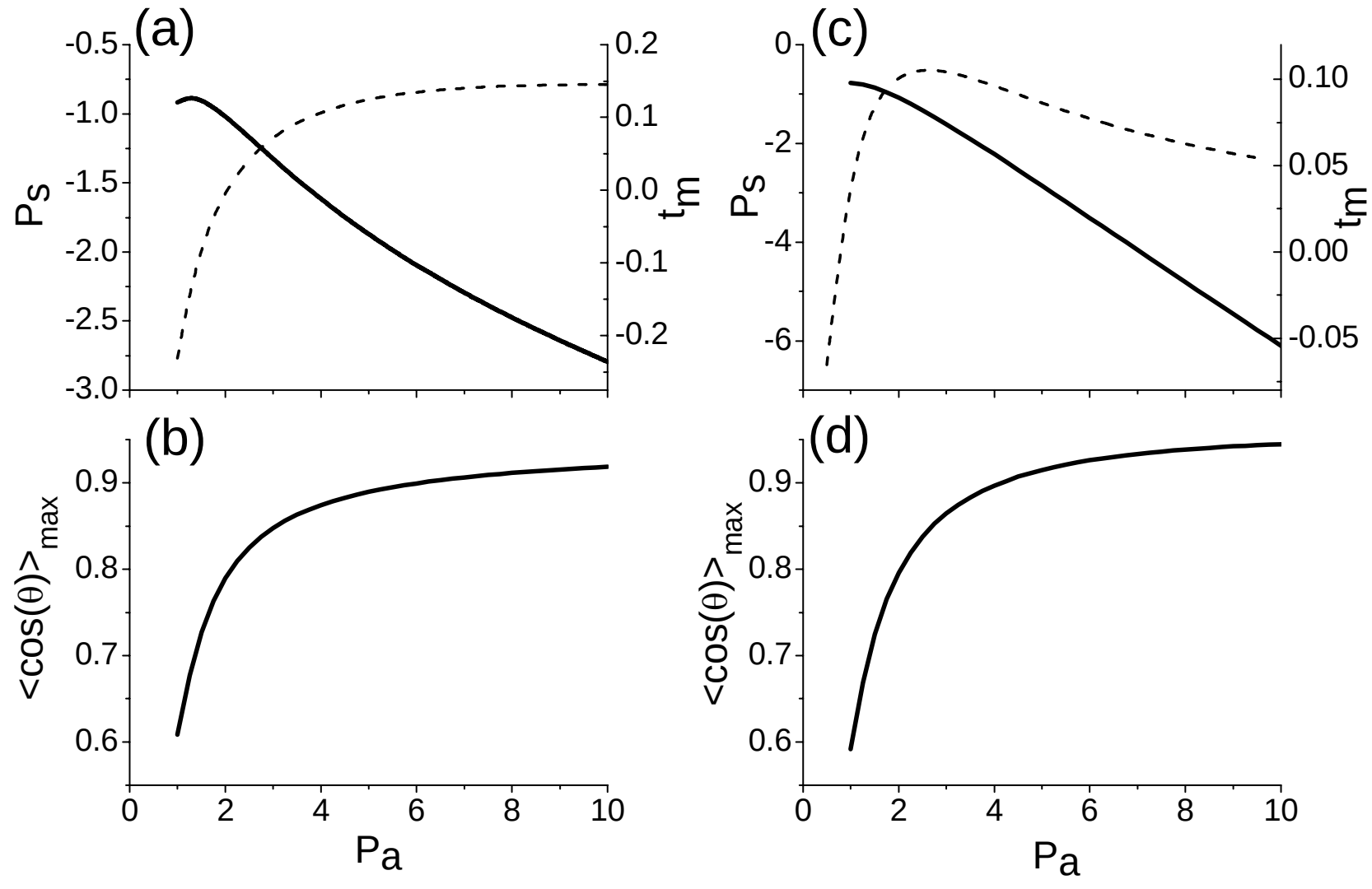

Figure 4. Quantum mechanical optimized results. The left column (a,b) corresponds to the mechanism of "orienting an anti-aligned state" (laser pulse fired before the HCP). The right column corresponds to the mechanism of "correcting the rotational velocity aberration" (inverse order of pulses). Upper panels (a,c) display the optimal strength of the anti-aligning pulse (solid lines) and delay between pulses (dashed lines) as a function of the HCP strength. Lower panels (b,d) present the highest value of the post-pulse orientation factor vs the strength of the HCP.

[1] R. Velotta, N. Hay, M. B. Mason, M. Castillejo, J. P. Marangos, Phys. Rev. Lett. **87**, 183901 (2001); J. Itatani, D. Zeidler, J. Levesque, M. Spanner, D.M. Villeneuve, P.B. Corkum, Phys. Rev. Lett. **94,** 123902 (2005).

[2] R. A. Bartels, T. C. Weinacht, N. Wagner, M. Baertschy, C. H. Greene, M. M. Murnane, H. C. Kapteyn, Phys. Rev. Lett., **88** 13903 (2002); V. Kalosha, M. Spanner, J. Herrmann, and M. Ivanov, Phys. Rev. Lett., **88**, 103901 (2002).

[3] R. J. Gordon, L. Zhu, W. A. Schroeder, T. Seideman, J. Appl. Phys. 94 669 (2003).

[4] J.J. Larsen, I. Wendt-Larsen, H. Stapelfeldt, Phys. Rev. Lett. **83,** 1123 (1999); M. Tsubouchi, B. J. Whitaker, L. Wang, H. Kohguchi, T. Suzuki, Phys. Rev. Lett. **86,** 4500 (2001); I. V. Litvinyuk, Kevin F. Lee, P. W. Dooley, D. M. Rayner, D. M. Villeneuve, P. B. Corkum, Phys. Rev. Lett. **90,** 233003 (2003).

[5] E.A. Shapiro, M. Spanner, M.Y. Ivanov, Phys. Rev. Lett., **91**, 237901 (2003); J. Mod. Optics, **52**, 897 (2005); K.F. Lee, D.M. Villeneuve, P.B. Corkum, E.A. Shapiro, Phys. Rev. Lett., **93**, 233601 (2004);

[6] J. P. Heritage, T. K. Gustafson, and C. H. Lin, Phys. Rev. Lett. **34**, 1299 (1975); J. Ortigoso, M. Rodríguez, M. Gupta, B. Friedrich, J. Chem. Phys. **110**, 3870 (1999); T. Seideman, Phys. Rev. Lett. **83** 4971 (1999).

[7] F. Rosca-Pruna and M. J. J. Vrakking, Phys. Rev. Lett. **87,** 153902 (2001).

[8] For a recent review on quantum revivals, see R. W. Robinett, Phys. Rep. **392**, 1 (2004).

[9] B. Friedrich and D. Herschbach, J. Chem. Phys. **111**, 6157 (1999); J. Phys. Chem. A **103**, 10 280 (1999); H. Sakai *et al.*, Phys. Rev. Lett. **90**, 083001 (2003).

[10] E. Charron, A. Giusti-Suzor, F.H. Mies, Phys. Rev. A **49,** R641 (1994); C. M. Dion, A. D. Bandrauk, O. Atabek, A. Keller, H. Umeda, and Y. Fujimura, Chem. Phys. Lett. **302**, 215 (1999); S. Guérin, L. P. Yatsenko, H. R. Jauslin, O. Faucher, and B. Lavorel, Phys. Rev. Lett. **88**, 233601 (2002); H. Ohmura and T. Nakanaga, J. Chem.Phys. **120**, 5176 (2004).

[11] C. M. Dion, A. Keller, and O. Atabek, Eur. Phys. J. D **14**, 249 (2001); M. Machholm and N. E. Henriksen, Phys. Rev. Lett. **87**, 193001 (2001).

[12] I. Sh. Averbukh and R. Arvieu, Phys. Rev. Lett. **87**, 163 601 (2001).

[13] M. Leibscher, I. S. Averbukh, and H. Rabitz, Phys. Rev. Lett. **90**, 213001 (2003); Phys. Rev. A **69**, 013402 (2004).

[14] D. Sugny, A. Keller, O. Atabek, D. Daems, C. M. Dion, S. Guérin, and H. R. Jauslin, Phys. Rev. A **69**, 033402 (2004).

[15] K. F. Lee, I. V. Litvinyuk, P. W. Dooley, M. Spanner, D. M. Villeneuve, and P. B. Corkum, J. Phys. B: At., Mol. Opt. Phys. **37**, L43 (2004).

[16] C. Z. Bisgaard, M. D. Poulsen, E. Péronne, S. S. Viftrup, and H. Stapelfeldt, Phys. Rev. Lett. **92**, 173004 (2004).

[17] H. Stapelfeldt, T. Seideman, Rev. Mod. Phys. 75 (2003) 543.

[18] T. Pluta, Mol. Physics, **99**, 1535 (2001).

[19] D. Daems, S. Guérin, D. Sugny, and H. R. Jauslin, Phys. Rev. Lett. **94**, 153003 (2005)

[20] A. Matos-Abiague and J. Berakdar, Phys. Rev. A **68**, 063411 (2003)

[21] J. Ortigoso, Phys. Rev. Lett. **93**, 073001 (2004)